\documentclass[sigconf]{acmart}
\pdfoutput=1

\AtBeginDocument{%
  }

\copyrightyear{2023}
\acmYear{2023}
\setcopyright{acmlicensed}
\acmConference[APNET 2023]{7th Asia-Pacific Workshop on Networking}{June 29--30, 2023}{Hong Kong, China}
\acmBooktitle{7th Asia-Pacific Workshop on Networking (APNET 2023), June 29--30, 2023, Hong Kong, China}
\acmPrice{15.00}
\acmDOI{10.1145/3600061.3600069}
\acmISBN{979-8-4007-0782-7/23/06}

\begin{document}

%%
%% The "title" command has an optional parameter,
%% allowing the author to define a "short title" to be used in page headers.
% \title{Bamboo: Improving Training Efficiency /Fast Training for Real-Time Video Streaming via Multi-User Online Transfer Learning}
\title{Bamboo: Boosting Training Efficiency for Real-Time Video Streaming via Online Grouped Federated Transfer Learning}
% \title{Bamboo: Boosting Training Efficiency for Real-Time Video Streaming via Multi-User Online Transfer Learning}
%\title{Bamboo: Boosting Training Efficiency for Real-Time Video Streaming via Online Grouped Asynchronous Distributed Transfer Learning}

%%
%% The "author" command and its associated commands are used to define
%% the authors and their affiliations.
%% Of note is the shared affiliation of the first two authors, and the
%% "authornote" and "authornotemark" commands
%% used to denote shared contribution to the research.
\author{Qianyuan Zheng}
\affiliation{%
  \institution{Nanjing University}
 % \streetaddress{1 Th{\o}rv{\"a}ld Circle}
   \city{Nanjing}
   \country{China}
  }
\email{qianyuanzheng@smail.nju.edu.cn}

\author{Hao Chen}
\authornote{Corresponding author: Hao Chen. This work is partially supported by the National Natural Science Foundation of China (62101241), Jiangsu Provincial Double-Innovation Doctor Program
(JSSCBS20210001).}
\affiliation{%
  \institution{Nanjing University}
 % \streetaddress{1 Th{\o}rv{\"a}ld Circle}
   \city{Nanjing}
    \country{China}
  }
\email{chenhao1210@nju.edu.cn}
\author{Zhan Ma}
\affiliation{%
  \institution{Nanjing University}
 % \streetaddress{1 Th{\o}rv{\"a}ld Circle}
   \city{Nanjing}
   \country{China}
  }
\email{mazhan@nju.edu.cn}

%%
%% By default, the full list of authors will be used in the page
%% headers. Often, this list is too long, and will overlap
%% other information printed in the page headers. This command allows
%% the author to define a more concise list
%% of authors' names for this purpose.
\renewcommand{\shortauthors}{Q. Zheng et al.}

%%
%% The abstract is a short summary of the work to be presented in the
%% article.
% \begin{abstract}
% Most of the learning-based algorithms for bitrate adaptation are limited to offline learning, which inevitably suffers from the simulation-to-reality gap. Online learning can better adapt to dynamic real-time communication scenes but still face the challenge of lengthy training convergence time. In this paper, we propose a novel online grouped federated transfer learning framework named \textsf{Bamboo} to accelerate training efficiency. The preliminary experiments validate that our method remarkably improves online training efficiency by up to 302\% compared to other reinforcement learning algorithms in various network conditions while ensuring the quality of experience (QoE) of real-time video streaming.
% \end{abstract}

%%
%% The code below is generated by the tool at http://dl.acm.org/ccs.cfm.
%% Please copy and paste the code instead of the example below.
%%


%%
%% Keywords. The author(s) should pick words that accurately describe
%% the work being presented. Separate the keywords with commas.
% \keywords{real-time video streaming, adaptive bitrate, transfer learning}

%%
%% This command processes the author and affiliation and title
%% information and builds the first part of the formatted document.
\maketitle
\vspace{-0.8em}
\section{Introduction}
\vspace{-0.2em}
Recently, learning-based bitrate adaptation algorithms have emerged, eliminating the reliance on handcrafted rules. For example, they utilize deep reinforcement learning (RL) to train their learning agents for the generation of proper bitrate adaptation policies in a simple simulation environment.
%using historical traces to radically simulate the live video streaming experience for the generation of proper bitrate adaptation policies.
%, which makes them better adapt to dynamic real-time communication scenes.

Existing works primarily involve offline and online learning methods.  Most of them are limited to the offline mode, applying the ``learning offline, running online'' strategy~\cite{onrl}, which inevitably suffers from the simulation-to-reality gap.
%. The learning models are trained in simulators and then deployed in real networks, which inevitably suffer from the simulation-to-reality gap~\cite{concerto}.
%, resulting in serious performance degradation when coping with the real internet. 
Compared with offline learning, online learning supports the training along with video streaming service, continuously refining RL models in response to new environments instead of relying on pre-trained models. %OnRL~\cite{onrl} designs an online reinforcement learning-based adaptation framework to close the simulation-to-reality gap, further improving the quality of experience (QoE) of real-time mobile video sessions.
% A typical user's real-time video streaming session will not last more than one hour~、cite{report}.
Although online learning can better adapt to dynamic scenes, it faces the slow training convergence challenge. On the one hand, acquiring the training data is relatively slow as environmental observations are not available for collection until the completion of a video streaming session. In this case, only one actual learning agent can be used for online learning at a time. On the other hand, it is impractical to wait for the online model to be fully trained before making decisions, considering that a real-time video streaming session enforces the prompt response~\cite{report}.

%\textcolor{red}{Furthermore, algorithms without prior experience rely on trial-and-error explorations particularly in the early stage of training, resulting in significant time wastage. On the one hand,  acquiring users' training data is relatively slow as the collection of corresponding observations must wait for the completion of each video streaming session. In this case, only one actual learning agent can be used for online learning at a time. On the other hand, it is impractical to wait for the online model to be fully trained before making decisions, considering that a real-time video streaming session enforces the prompt response~\cite{report}. }%We optimize the model's convergence speed while ensuring communication performance, hoping to conduct online training within a user's real-time video session.

To this end, this paper proposes \textsf{Bamboo}, a novel online grouped federated transfer learning framework for real-time video streaming to boost training efficiency. \textsf{Bamboo} utilizes user grouping for intra-group federated learning and mitigates RL's trial-and-error impacts through online transfer. As a result, \textsf{Bamboo} can complete the online training within a user's real-time video session.

We implement \textsf{Bamboo} on a WebRTC-based real-time video conferencing testbed\footnote{We develop the testbed based on an open-source WebRTC framework which is available at https://github.com/yuanrongxi/razor.} which uses the Linux traffic control (TC) tool to control the network condition. Our experimental results show that \textsf{Bamboo} remarkably improves online training efficiency by up to 302\% compared to other reinforcement learning algorithms across various network conditions while ensuring the quality of experience (QoE) of real-time video streaming.
%Online transfer learning is carried out in the real client video player after a pre-trained offline model, and online fine-tuning is additionally activated to improve further the ability to generalize to the network conditions and video source characteristics that have never been met before. Multi-users are cooperatively trained in groups according to network types and transportation modes to accelerate training data acquisition. Multi-users regularly share model data with the central server, thus reducing the training time and computing resources required by each user.
% We implement Bamboo on a WebRTC~\cite{webrtc} based real-time video conferencing testbed Echo\footnote{We develop Echo based on an open-source WebRTC framework which is available at https://github.com/yuanrongxi/razor.} which uses Linux tc~\cite{tc} to control the network condition. Our preliminary evaluation demonstrates a performance training efficiency improvement of up to xxx\% over other reinforcement learning bitrate adaptation algorithms in a variety of network environments while ensuring the QoE of real-time video streaming.
%Our preliminary evaluation demonstrates the performance training efficiency has tripled compared to other reinforcement learning bitrate adaptation algorithms in various network conditions while ensuring the QoE of real-time video streaming.
%%%%%%%%%%%%%%%%%%%%%%%%%%%%%%%%%%%%%%%%%%%%%%%%%%%%%%%%%%%%%%%%%%%%%%%%
\vspace{-0.8em}
\section{Bamboo Design}
\vspace{-0.2em}
\begin{figure}[t]
    \setlength{\abovecaptionskip}{0.cm}
    \centering
    \includegraphics[width=0.95\linewidth]{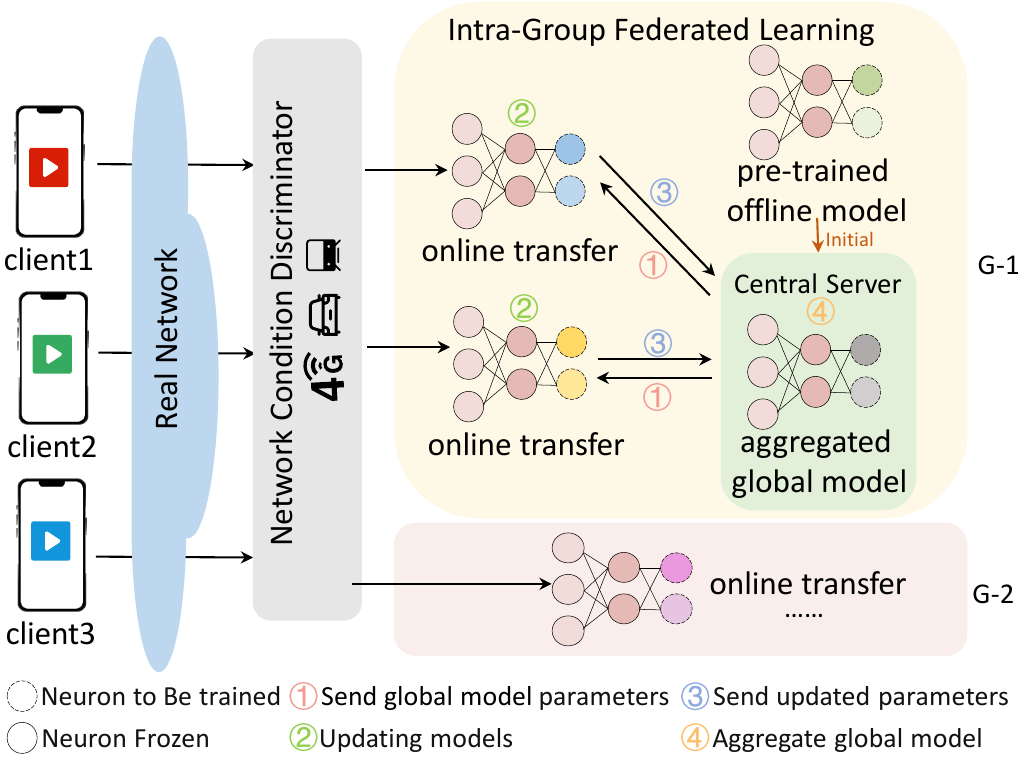}
    \caption{\textsf{Bamboo} utilizes the discriminator to group users and then performs grouped federated learning within each group. Users conduct online transfer learning.}
    \label{fig:Bamboo}
\vspace{-0.62cm}
\end{figure}
%%%%%%%%%%%%%%%%%%%%%%%%%%%%%%%%%%%%%%%
% We illustrate the workflow of Bamboo’s adaptive learning
% aggregation in xxxx.
%%%%%%%%%%%%%%%%%%%%%%%%%%%%%%%%%%%
Figure~\ref{fig:Bamboo} gives an overview of \textsf{Bamboo}. We first design a dynamic network condition discriminator to identify each user's network type and transportation mode. Especially, the network type (e.g., 3G, 4G, and WiFi) can be directly obtained from internet service providers (ISP)~\cite{legato}, and the transportation mode (e.g., car, bus, ferry, and train) can be detected using mobile sensors (e.g., GPS, accelerometers or microphones)~\cite{transport}. Users are classified into 12 groups (labeled as $G$, which can be extended to the more general group classification) as depicted in Table~\ref{tab:network_classifies}. Users within the same group engage in intra-group federated learning. 

{\bf Intra-Group Federated Learning.} Without loss of generality, we assume two users and one central server are in one group, which can be extended to more general cases. The framework primarily comprises four procedures. Firstly, offline training involves training a well-generalized fixed neural network model using extensive datasets in a simulator, which is then deployed on the central server. Secondly, the pre-trained offline model is distributed to users, enabling them to train their own models using local network conditions. Thirdly, each user locally updates the model parameters and periodically shares them with the central server. The central server aggregates these parameters to construct an aggregated global model of the group. Finally, users can train personalized models by incorporating the central server model with their own previous models. The aggregated global model undergoes iterative updates through interactive exchanges with the users in a cyclic iterative process. This accelerates batch writing of user data and facilitates RL model training within the same group.
%%%%%%%%%%%%%%%%%%%
\begin{table}[t]
\setlength{\abovecaptionskip}{0.cm}
\caption{\textsf{Bamboo} classifies network conditions into 12 groups.}
%\vspace{-1.0em}
\label{tab:network_classifies}
\begin{tabular}{ll|ll|ll}
\hline
$G$-1:\quad &3G \& foot      & $G$-5:\quad &4G \& foot & $G$-9:\quad &WIFI \& foot\\ \hline
$G$-2:\quad &3G \& car      & $G$-6:\quad &4G \& car & $G$-10:\quad &WIFI \& car\\ \hline
$G$-3:\quad &3G \& ferry      & $G$-7:\quad &4G \& ferry & $G$-11:\quad &WIFI \& ferry\\ \hline
$G$-4:\quad &3G \&  train     & $G$-8:\quad &4G \& train & $G$-12:\quad &WIFI \& train\\ \hline
\end{tabular}
\vspace{-2.1em}
\end{table}

%%%%%%%%%%%%%%%%
Specifically, the federal learning method we adopt enhances the A3C~\cite{a3c} algorithm to synchronously optimize the neural network of each user by averaging the optimization of neural network gradients within the same group. Users dynamically interact with the real-time video streaming environment, undergo training, and make decisions. The integration of federated learning techniques not only mitigates communication overhead resulting from data transmission but also simultaneously enhances model accuracy and generalization performance.

% {\bf Online transfer learning.} 
{\bf Efficient Online Transfer.} In the final step, to mitigate the discrepancy between the offline pre-trained model and the user's online model caused by the simulation-to-reality gap, we use online transfer training to fine-tune the pre-trained model to better adapt to the real network conditions and video content characteristics. %, aiming to optimize the bitrate adaptation algorithm. %In model transfer, we think that the convolution layers aim at extracting low-level features about activity recognition. Thus we keep these layers frozen, so we do not update their parameters in backpropagation. As for the fully connected layers, since they are higher level, we believe they focus on learning specific features for the task and user. Therefore, we update their parameters during training.
In model transfer, the low-level layers for common feature extraction are frozen. In contrast, the parameters of high-level task-specific layers are updated to adapt to the new network environment. 

Through efficient offline pre-training,  the basic knowledge of the real-time video streaming environment is learned and the scarcity of real samples is alleviated. In the following online training, transfer learning is leveraged to fully utilize past experiences for mitigating early-stage ``trial-and-error'' and expediting the training process for new network environments, which significantly benefits computational resource savings without degrading model generalization.

Overall, the dynamic network condition discriminator periodically identifies the group to which a user belongs, and the user continuously interacts with the central server within the same group to exchange the model parameters. Such flexible grouping and iterative training endow \textsf{Bamboo} to handle instantaneous changes in network conditions and enhance its generalization performance.
\vspace{-0.8em}
\section{EVALUATION \& FUTURE WORK}
\vspace{-0.2em}
We built a WebRTC-based real-time video conferencing testbed to evaluate \textsf{Bamboo}. We created a network trace corpus from HSDPA~\cite{3g}, NYU~\cite{nyu}, FCC~\cite{fcc}, and used the Linux TC tool to reproduce the real-world network dynamics recorded in these traces. The corpus was randomly divided into training and testing datasets, with 80\% of the data used for training \textsf{Bamboo} and 20\% for testing all algorithms by default. Within the training set, 80\% of data were used to train pre-trained offline models, while the remaining 20\% were used to train online fine-tuning models. We compare \textsf{Bamboo} to the following schemes, which represent the state-of-the-art in bitrate adaptation: (a) ARS~\cite{ars}, a learning-based bitrate adaptation algorithm with its model trained offline; (b) OnRL~\cite{onrl}, which employs online training from scratch to improve its model performance; (c) \textsf{Bamboo}-transfer, a variant of \textsf{Bamboo} with transfer learning enabled but intra-group federated learning disabled. %, (d) \textsf{Bamboo}-transfer with intra-group federated Learning, all modules putting together (\textsf{Bamboo}, assuming two users and one central server are in one group).
Note that we use the same neural network architecture for these baselines as \textsf{Bamboo} for a fair comparison.

%%%%%%%%%%%%%%%%%%%%%%%%%%%%%%%%
\begin{figure}[t]
%\vspace{-12pt}
\centering
\begin{minipage}[h]{0.46\linewidth}
\centering
\setlength{\abovecaptionskip}{0pt}
\includegraphics[width=\linewidth]{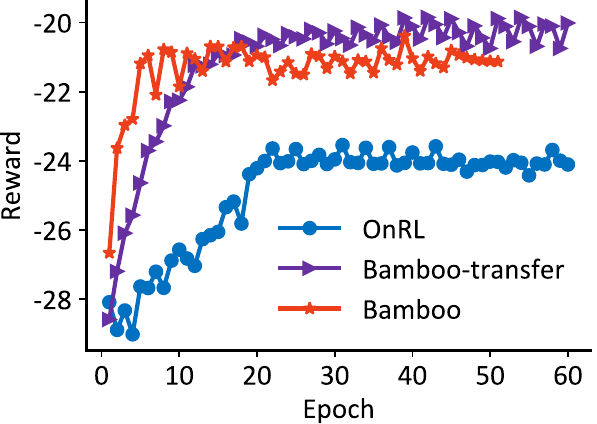}
\caption{Rewards gained over epochs in the training.}
\label{fig:log}
\end{minipage}
%%%%%%%%%%%%%%%%%%%%%%%%%%%%%%%%
\hspace{0.1cm}
\begin{minipage}[h]{0.48\linewidth}
%\vspace{-3pt}
\centering
\setlength{\abovecaptionskip}{-5pt}
\includegraphics[width=\linewidth]{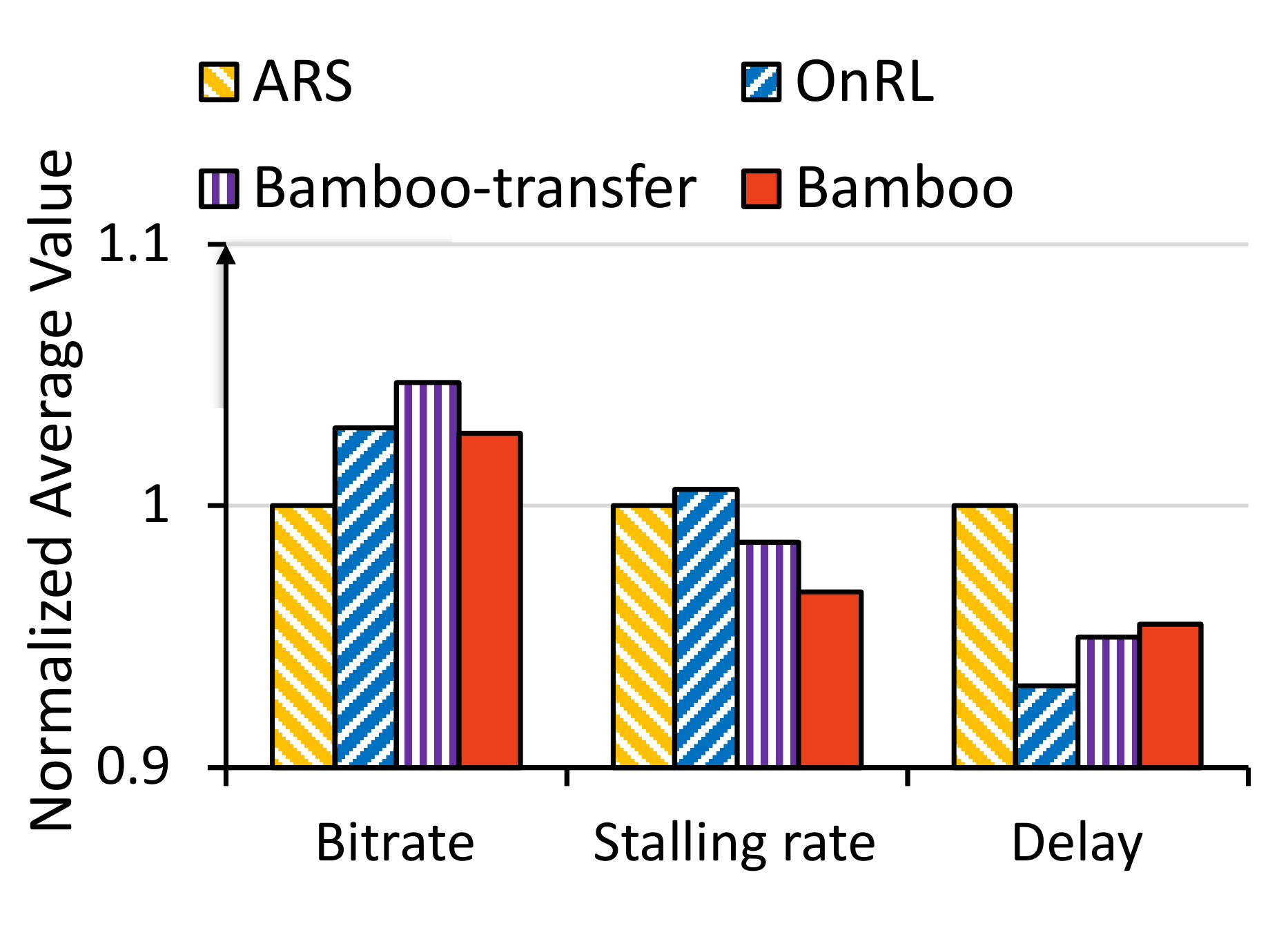}
\caption{Performance evaluation on each QoE metric.}
\label{fig:CDF}
\end{minipage}
\vspace{-14pt}
\end{figure}

%%%%%%%%%%%%%%%%%%%%%%%%%%%%%%%%%%%%%%%%%%%%%%%%%%%%%%%%%%%%%%%%%%%%%%%
\begin{table}[t]
\setlength{\abovecaptionskip}{0.cm}
\setlength{\belowcaptionskip}{0cm}
\caption{Average convergence time in hours to train an online model using different schemes on various networks.}
\label{tab: convergence time}
\begin{tabular}{c|c|c|c}
\hline
   & 3G & 4G & WIFI \\ \hline
OnRL & 2.54 & 2.65 & 2.57 \\ \hline
\textsf{Bamboo}-transfer & 1.56 & 1.28 & 1.51 \\ \hline
\bf{\textsf{Bamboo}} & \bf{0.80} & \bf{0.53} & \bf{0.60} \\ \hline
\end{tabular}
\vspace{-2em}
\end{table}
%%%%%%%%%%%%%%%%%%%%%%%%%%%%%%%%%%%%%%%%%%%%%%%%%%%%%%%%%%%%%%%%%%%%%%%
Table~\ref{tab: convergence time} and Figure~\ref{fig:log} illustrate the training efficiency of \textsf{Bamboo}. We evaluate the performance of \textsf{Bamboo} and other methods under the same test trace lasting for 300s and plot the normalized average value of QoE metrics (e.g., bitrate, stalling rate, and delay) in Figure~\ref{fig:CDF}. We observe that \textsf{Bamboo}'s online transfer and grouped federated training improve the online model training efficiency by 43.9\% and 55.6\%, respectively. Furthermore, when compared to the ARS anchor across all trace sets, these mechanisms increase the bitrate by 0.7\%-3.7\% and reduce the stalling rate by 2.3\%-2.9\% as well as delay by 2.3\%-4.7\%. These findings demonstrate \textsf{Bamboo}'s proficiency in knowledge transfer and its generalization ability by mitigating RL's trial-and-error impacts and expediting training data acquisition. Consequently, \textsf{Bamboo} outperforms existing approaches by 302\% in terms of average convergence time and ensures a satisfactory QoE in real-time video streaming.

In the future, we will consider more practical issues (e.g., robust hybrid learning mechanism, determining whether the online training can be terminated) and conduct more testbed experiments.

%%%%%%%%%%%%%%%%%%%%%%%%%%%%%%%%%%%%%%%%%%%%%%%%%%%%%%%%%%%%%%%%%%%%%%%% CONCLUSION AND FUTURE WORK
% \section{CONCLUSION AND FUTURE WORK}

% We propose \textsf{Bamboo} to improve training efficiency based on online grouped federated transfer learning. Preliminary experiments validate that our method is effective. Our future work mainly focuses on robust hybrid learning mechanisms to ensure the performance reliability of online learning, how to determine the online training can be terminated, and work with our current work jointly to further improve training efficiency.
\vspace{-0.8em}
\bibliographystyle{ACM-Reference-Format}
\bibliography{bamboo}
% \bibliography{acmart}

\end{document}